\def\Title{Universally valid reformulation of the Heisenberg
uncertainty principle on noise and disturbance in measurement}
\def\Author{Masanao Ozawa}
\documentstyle[twocolumn,aps]{revtex}  

  \newcommand{\beq}{\begin{equation}}
  \newcommand{\eeq}{\end{equation}}
  \newcommand{\beql}[1]{\begin{equation}\label{eq:#1}}
  \newcommand{\beqa}{\begin{eqnarray}}
  \newcommand{\eeqa}{\end{eqnarray}}
  \newcommand{\beqas}{\begin{eqnarray*}}
  \newcommand{\eeqas}{\end{eqnarray*}}
  
  \newcommand{\bA}{{\bf A}}
  \newcommand{\bB}{{\bf B}}
  \newcommand{\bC}{{\bf C}}

  \newcommand{\bP}{{\bf P}}
  
  \newcommand{\bS}{{\bf S}}

  \newcommand{\bx}{{\bf x}}
  \newcommand{\by}{{\bf y}}

  \newcommand{\da}{\dagger}
  
  \newcommand{\ep}{\epsilon}
  \newcommand{\et}{\eta}
  \newcommand{\ga}{\gamma}

  \newcommand{\nn}{\nonumber}
  \newcommand{\om}{\omega}
  
  \newcommand{\ps}{\psi} 
  
  \newcommand{\si}{\sigma} 
  
  \newcommand{\th}{\theta} 
  
  \newcommand{\ve}{\varepsilon}

  \newcommand{\De}{\Delta}                                          
  
  \newcommand{\Eq}[1]{Eq.~(\ref{eq:#1})}

  \newcommand{\eq}[1]{(\ref{eq:#1})}

\newcommand{\ket}[1]{|#1\rangle}

\newcommand{\bracket}[1]{\langle#1\rangle}
	
\newcommand{\A}{{\bf A}}
\newcommand{\In}{{\rm in}}
\newcommand{\Out}{{\rm out}}

\begin{document}
\draft
\title{\Title}
\author{\Author}
\address{Graduate School of Information Sciences,
T\^{o}hoku University, Aoba-ku, Sendai,  980-8579, Japan}
\maketitle
\begin{abstract}
The Heisenberg uncertainty principle${\cite{Hei81}}$
states that the product of the noise in a position
measurement and the momentum disturbance caused by that measurement
should be no less than the limit set by Planck's constant, 
${\hbar/2}$, as demonstrated by Heisenberg's thought experiment
using a ${\ga}$-ray microscope${\cite{Hei30}}$.
Here I show that this common assumption is false:
a universally valid trade-off relation between the noise and the disturbance
has an additional correlation term, which is redundant when the intervention
brought by the measurement is independent of the measured object, 
but which allows the noise-disturbance product much below Planck's
constant when the intervention is dependent.
A model of measuring interaction with dependent intervention shows that 
Heisenberg's lower bound for the noise-disturbance product is violated 
even by a nearly nondisturbing, precise position measuring instrument.   
An experimental implementation is also proposed to realize the above 
model in the context of optical quadrature measurement with currently
available linear optical devices. 
\end{abstract}

	

\section{Three formulations of uncertainty principle}

The uncertainty principle has been known as one of the
most fundamental principles in quantum mechanics.
However, there is still ambiguity in formulation;
we have at least three different formulations of
the ``uncertainty principle''.

The {\em Robertson uncertainty relation}
is generally formulated as the relation 
\beql{URSP}
\si(A,\ps)\si(B,\ps)\ge\frac{|\bracket{\ps|[A,B]|\ps}|}{2}
\eeq
for any observables $A$, $B$ and any state $\ps$,
where the standard deviation $\si(X,\ps)$ of an observable $X$ in
state $\ps$ is defined by $\si(X,\ps)^{2}=\bracket{\ps|X^{2}|\ps}
-\bracket{\ps|X|\ps}^{2}$.
This relation was proven mathematically from fundamental
postulates of quantum mechanics${\cite{Ken27,Rob29}}$.  
Nevertheless, this relation describes the limitation on preparing 
microscopic objects but has no direct relevance to
the limitation of accuracy of measuring devices${\cite{Bal70}}$.

It is a common understanding that the uncertainty principle implies 
or is implied by a limitation on measuring a system without disturbing it
as a position measurement typically disturbs the momentum.
However, the limitation has eluded a correct quantitative expression
on the trade-off between noise and disturbance.
By the $\ga$-ray thought experiment, Heisenberg${\cite{Hei81,Hei30}}$ argued that 
the product of the noise in a position measurement and the momentum
disturbance caused by that measurement
should be no less than $\hbar/2$.
This relation is generally formulated as follows:  
For any apparatus $\bA$
to measure an observable $A$, the relation
\beql{URMD}
\ep(A,\ps,\bA)\et(B,\ps,\bA)\ge\frac{|\bracket{\ps|[A,B]|\ps}|}{2}
\eeq
holds for any input state $\ps$ and any observable $B$,
where $\ep(A,\ps,\bA)$ stands for the noise of the $A$ measurement
in state $\ps$ using apparatus $\bA$ and $\et(B,\ps,\bA)$ stands
for the disturbance of $B$ in state $\ps$ caused by apparatus $\bA$.
We refer to the above relation as the {\em Heisenberg noise-disturbance
uncertainty relation}.
We shall investigate the validity of this relation to solve such questions as:
When does this relation
hold and when does not?  What relation can be considered a universally
valid generalization of this relation?  How can we experimentally observe
the violation of this relation?

Closely related to the above relation, the {\em Heisenberg uncertainty
relation for joint measurements} is generally formulated as follows:  For
any apparatus $\bA$ with two outputs for the joint measurement of $A$
and
$B$,  the relation
\beql{URJM}
\ep(A,\ps,\bA)\ep(B,\ps,\bA)\ge\frac{|\bracket{\ps|[A,B]|\ps}|}{2}
\eeq
holds for any input state $\ps$,
where $\ep(X,\ps,\bA)$ stands for the noise of the $X$ measurement
in state $\ps$ using apparatus $\bA$ for $X=A,B$.
This relation was proven mathematically${\cite{AK65,AG88,91QU,Ish91}}$
under the {\em joint unbiasedness condition} 
requiring that the (experimental) mean values of the outcome $\bx$ of 
the $A$ measurement and the outcome $\by$ of the $B$
measurement should coincide with the (theoretical) mean values of
observables $A$ and $B$, respectively, on any input state $\ps$.  
It is a common opinion that currently available measuring devices satisfy this
relation${\cite{SH66,Yue82,YH86}}$. 

There is a logical relationship between the noise-disturbance relation,
\eq{URMD}, and the joint measurement relation, \eq{URJM}.
Suppose that the $A$ measurement using the apparatus $\bA$ is  
followed immediately by a measurement of the observable $B$ using a
noiseless measuring apparatus
$\bB$.  Then, combining the two apparatuses, we have a new apparatus
$\bC$ to jointly measure $A$ and $B$ on the input state of apparatus
$\bA$.  
Since $\bB$ carries out a noiseless measurement on its own input
state, the noise of $B$ measurement in the outcome of apparatus 
$\bC$ is equal to the $B$ disturbance caused by apparatus $\bA$. 
Thus, we have the relations
\begin{mathletters}\label{eq:MD-JM}
\beqa
\ep(A,\ps,\bA)&=&\ep(A,\ps,\bC),\\
\et(B,\ps,\bA)&=&\ep(B,\ps,\bC).
\eeqa
\end{mathletters}%
By applying the uncertainty relation for joint measurements
to the apparatus $\bC$, we can conclude that the noise-disturbance
relation, \eq{URMD},  holds if apparatus $\bC$ satisfies the joint
unbiasedness condition for observables $A$ and $B$.
However, there are two deficiencies of the above approach: (i) Even for
noiseless  measuring apparatus $\bA$ to measure $A$ one cannot ensure that the 
combined apparatus $\bC$ satisfies the joint unbiasedness condition.
(ii) The above argument does not give a universally valid 
trade-off relation between noise and disturbance.
Thus, we can conclude that the validity of the 
noise-disturbance relation, \eq{URMD}, cannot be reduced to
the current formulation of the Heisenberg uncertainty relation for joint
measurements, \eq{URJM}.

\section{Measurement noise and disturbance}

Now, let us analyze noise and disturbance in a most
general description of measurement${\cite{00MN,01OD,02CU}}$
in detail. 
Let ${\A}$ be a measuring apparatus with (macroscopic) output
variable $\bx$ to measure an observable $A$ of a quantum system
$\bS$. 
Then, apparatus ${\A}$ measures observable $A$
precisely if and only if the (experimental) probability distribution of ${\bf
x}$ coincides with the (theoretical) probability distribution of $A$; 
or rigorously the probability of the event that the output $\bx$ of 
apparatus $\A$ is in an interval $\De$
satisfies the {\em Born statistical formula (BSF)} for observable $A$,
i.e., 
\beql{010530a}
\Pr\{\bx\in\Delta\}=\bracket{\ps|E^{A}(\De)|\ps}
\eeq
on every input state $\ps$,
where $E^{A}(\De)$ stands for the spectral projection of $A$
corresponding to interval $\De$.
Otherwise, we consider apparatus
${\A}$ to measure observable $A$ with some noise.
In order to evaluate the noise, we need to describe
the measuring process.
The measuring interaction is supposed to turn on at a time $t$, the 
{time of measurement}, and to turn off at time $t+\De t$  between the
object $\bS$ and a quantum subsystem $\bP$, called the {\em probe}, of the (large)
apparatus ${\A}$.
Denote by $U$ the unitary operator representing the
time evolution of the composite system $\bS+\bP$ for 
the time interval $(t,t+\Delta t)$.   
We assume that the object and any part of the
apparatus do not interact  before $t$ nor after $t+\Delta t$.
At the time of measurement the object is supposed to
be in an arbitrary state $\ps$ and the probe is
in a fixed state $\xi$.
Thus, the composite system $\bS+\bP$ is in the state $\psi\otimes\xi$
at time $t$.  Just after the measuring interaction has turned off, 
the probe is subjected to a precise local measurement of
an observable $M$ of the probe, called the {\em probe observable},
and its output is recorded by the output variable $\bx$.  
In the Heisenberg picture with the original state $\psi\otimes\xi$ at time
$t$, we write 
$A^{\In}=A\otimes I$,
$M^{\In}=I\otimes M$,
$A^{\Out}=U^{\da}(A\otimes I)U$,
and $M^{\Out}=U^{\da}(I\otimes M)U$.
Since the output 
of this measurement is obtained by the precise measurement of 
observable $M$ at time $t+\De t$,  the probability 
distribution of the output variable $\bx$ is given by
\beql{B1}\label{eq:0328b}
\Pr\{\bx\in\De\}
=\bracket{E^{M^{\Out}}(\De)},
\eeq
where $\bracket{\cdots}$ stands for
$\bracket{\psi\otimes\xi|\cdots|\psi\otimes\xi}$ throughout this paper.  

The {\em noise} $\ep(A,\ps,\A)$, or $\ep(A)$ for short, of the $A$ 
measurement
on input state $\ps$ using apparatus ${\A}$ is defined as
the root-mean-square deviation of the experimental variable $M^{\Out}$
from the theoretical variable $A^{\In}$,  i.e.,  
\beql{noise}
\ep(A,\ps,\A)=\bracket{(M^{\Out}-A^{\In})^{2}}^{1/2}.
\eeq
Then, we can prove that {\em $\ep(A)=0$ on any input state $\ps$ if
and only if apparatus ${\A}$ measures observable
$A$ precisely.}
This ensures that ``precise apparatuses''
and ``noiseless apparatuses'' are equivalent.
Let $B$ be an observable of $\bS$.

The {\em disturbance} $\et(B,\ps,\A)$, or $\et(B)$ for short, 
of observable $B$ on input state $\ps$ caused by apparatus ${\A}$ 
is defined  as the root-mean-square of the change in the observable $B$ 
during the 
measuring interaction, i.e., 
\beql{disturbance}
\et(B,\ps,\A)=\bracket{(B^{\Out} -B^{\In})^{2}}^{1/2}.
\eeq
Then, we can prove that {\em $\et(B)=0$ on any input state $\ps$
if and only if apparatus ${\A}$ does not change the probability 
distribution of the observable $B$, i.e., 
$
\bracket{E^{B^{\In}}(\De)}=\bracket{E^{B^{\Out}}(\De)}
$
for every interval  $\De$ on any input state $\ps$}. 
Thus, apparatuses not disturbing the observable $B$ 
and apparatuses with zero disturbance, $\et(B)\equiv 0$, 
are equivalent notions.
It should be also noted that the above definitions of noise and
disturbance are consistent with the standard formulation
for the Heisenberg uncertainty relation for joint measurements,
\Eq{URJM}, with \Eq{MD-JM}.
Thus, the above definitions of noise and disturbance can be considered
standard.

\section{Universally valid uncertainty principle}

Under the above general definitions of noise and disturbance,
we can rigorously investigate the validity of the Heisenberg 
noise-disturbance uncertainty relation, \Eq{URMD}.
For this purpose,  we introduce the {\em noise operator} $N(A)$
and the {\em disturbance operator} $D(B)$ by
\beqa
M^{\Out}&=&A^{\In}+N(A),\label{eq:NO}\\
B^{\Out}&=&B^{\In}+D(B).\label{eq:DO}
\eeqa
Since $M$ and $B$ are observables in different systems, we have
$[M^{\Out},B^{\Out}]=0$, and hence we have
the following commutation relation for the noise operator
and the disturbance operator, 
\beqa
& &[N(A),D(B)]+[N(A),B^{\In}]+[A^{\In},D(B)]\nn\\
& &\qquad=-[A^{\In},B^{\In}].
\eeqa
Taking the modulus of means of the both sides
 and applying the triangular inequality, we have
\beqa
& &
|\bracket{[N(A),D(B)]}|+|\bracket{[N(A),B^{\In}]}
+\bracket{[A^{\In},D(B)]}|
\nn\\
& &\qquad
\ge|\bracket{\ps|[A,B]|\ps}|.
\eeqa
Since the variance is not greater than the mean square,  we have 
\beqas
\ep(A,\ps,\bA)&\ge&\si(N(A),\ps\otimes\si),\\
\et(B,\ps,\bA)&\ge&\si(D(B),\ps\otimes\si),
\eeqas
and hence by the Heisenberg uncertainty principle,  \Eq{URSP}, we have
\beq
\ep(A)\et(B)\ge\frac{|\bracket{[N(A),D(B)]}|}{2}.
\eeq
Thus, we obtain the {\em universally valid noise-disturbance uncertainty
relation} for the pair $(A,B)$,
\beqa\label{eq:UVUR}
& &\ep(A)\et(B)
+\frac{|\bracket{[N(A),B^{\In}]}+\bracket{[A^{\In},D(B)]}|}{2}\nn\\
&  &\qquad\ge\frac{|\bracket{\ps|[A,B]|\ps}|}{2}.
\eeqa
The above relation shows that the Heisenberg noise-disturbance
uncertainty relation,
\Eq{URMD}, holds if the correlation term
$|\bracket{[N(A),B^{\In}]}+\bracket{[A^{\In},D(B)]}|$ vanishes.
In order to characterize a class of measurements satisfying
\Eq{URMD}, we introduce the following  definition.
The measuring interaction is said to be
of {\em independent intervention} for the pair $(A,B)$ if the noise
and the disturbance are independent of the object system; or precisely if
there is observables $N$ and $D$ of probe $\bP$ such that
$N(A)=1\otimes N$ and $D(B)=1\otimes D$. In this case, we have
$[N(A),B^{\In}]=[A^{\In},D(B)]=0$. 
Therefore, we conclude that {\em if
the measuring interaction $U$ is of independent intervention for the pair
$(A,B)$,  the Heisenberg noise-disturbance uncertainty relation for 
the pair $(A,B)$, \Eq{URMD}, holds for any object state $\ps$ and any
probe state $\xi$.} 
The above conclusion was previously suggested in part 
by Braginsky and
Khalili${\cite{BK92}}$ with a limited justification and now fully
justified. 

The universally valid uncertainty relation shows that for measurements
of dependent intervention, the lower bound of the noise-disturbance
product depends on the pre-measurement uncertainties 
(standard deviations) of $A$ and $B$. 
In order to obtain the trade-off among the noise
$\ep(A)$, the disturbance $\et(B)$, and the pre-measurement uncertainties
$\si(A)$ and $\si(B)$, we apply the Heisenberg uncertainty principle,
\Eq{URSP}, to all terms in the left-hand-side of the universally
valid noise-disturbance uncertainty relation, \Eq{UVUR}.  Then, we now
obtain the {\em generalized noise-disturbance uncertainty relation},
\beql{GNDUR}
\ep(A)\et(B)+\ep(A)\si(B)+\si(A)\et(B)
\ge
\frac{|\bracket{\ps|[A,B]|\ps}|}{2}.
\eeq
The above relation holds for any apparatus $\bA$
(specified by any probe state $\xi$, any unitary operator $U$,
and any probe observable $M$), any observables $A,B$,
and any input state $\ps$, and hence ultimately generalizes the 
Heisenberg noise-disturbance uncertainty relation, \Eq{URMD},
to arbitrary measurements.

Under the finite energy constraint, i.e., $\si(Q), \si(P)<\infty$,
the above relation excludes the possibility of having both
$\ep(Q)=0$ and $\et(P)=0$.
However,
$\ep(Q)=0$ is possible with $\si(Q)\et(P)\ge\hbar/2$
and $\et(P)=0$ is possible with $\ep(Q)\si(P)\ge\hbar/2$.
In particular, even the case where $\ep(Q)=0$ and $\et(P)<\ve$ 
with arbitrarily small $\ve$ is possible for some input state with 
$\si(Q)>\hbar/2\ve$,
and also the case where $\et(P)=0$ and $\ep(Q)<\ve$ is
possible for some input state with $\si(P)>\hbar/2\ve$.
Such extreme cases occur in compensation for large uncertainties
in the input state, while
in the minimum uncertainty state with
$\si(Q)=\si(P)=(\hbar/2)^{1/2}$, we have 
\beq
\ep(Q)\et(P)+\sqrt{\frac{\hbar}{2}}[\ep(Q)+\et(P)]\ge\frac{\hbar}{2}.
\eeq
Even in this case, it is allowed to have $\ep(Q)\et(P)=0$ with 
$\ep(Q)=0$ and $\et(P)\ge(\hbar/2)^{1/2}$ or with $\et(P)=0$ and 
$\ep(Q)\ge(\hbar/2)^{1/2}$.

\section{Violation of the Heisenberg inequality}

Now let us consider the problem as to whether  
one can implement, under the current experimental technique,
a good measuring apparatus with small noise-disturbance
product beyond the original Heisenberg lower bound. 
The controversy${\cite{Mad88}}$ on the sensitivity limit of gravitational
wave detectors suggested that the Heisenberg noise-disturbance
uncertainty relation is not universally valid.
In fact, based on the Heisenberg noise-disturbance uncertainty relation,   
Braginsky, Caves, and others${\cite{Bra74,CTDSZ80}}$
claimed that there is a sensitivity limit, called the
standard quantum limit (SQL),  for monitoring the free-mass position
that leads to a quantum mechanical sensitivity limit on
interferometer type gravitational wave detectors.
However, Yuen${\cite{Yue83}}$ proposed the idea of  ``contractive state
measurements'' to break the SQL and Ozawa${\cite{88MS,89RS}}$ found an
explicit Hamiltonian realization of a contractive state measurement that
breaks the SQL.  
Consequently, the above measuring interaction violates the Heisenberg
noise-disturbance uncertainty relation.  Direct computaions on
the position-measuring-niose and momentum-disturbance was given in
\cite{02KB5E} showing the violation of the Heisenberg noise-disturbance
uncertainty relation.

  In what follows, modifying the above interaction in the
context of optical quadrature measurement, I will show that the small
noise-disturbance product can be achieved beyond Heisenberg's lower bound 
by an apparatus carrying out a precise and nearly non-disturbing 
quadrature measurement with currently available linear optical devices. 

\section{Backaction evading quadrature amplifiers}

Consider the case where the system $\bS$ and the probe $\bP$ are
two optical modes represented by annihilation operators $a$ and $b$,
respectively.
The quadrature component field operators 
$X_{a}, Y_{a}, X_{b}, Y_{b}$
are self-adjoint operators satisfying $a=X_{a}+i Y_{a}$ and
$b=X_{b}+i Y_{b}$,
for which we have 
$[X_{a},Y_{a}]={i}/{2}$ and 
$[X_{b},Y_{b}]={i}/{2}$.

A measuring interaction $U$ on $\bS+\bP$ is called a 
{\em back-action evading (BAE)  quadrature amplifier}${\cite{Yur85}}$ with
gain $G$ if we have
\begin{mathletters}\label{eq:712b}
\beqa
X_{a}^{\Out}&=&X_{a}^{\In},\\
X_{b}^{\Out}&=&X_{b}^{\In}+GX_{a}^{\In},\\
Y_{a}^{\Out}&=&Y_{a}^{\In}-GY_{b}^{\In},\\
Y_{b}^{\Out}&=&Y_{b}^{\In}.
\eeqa
\end{mathletters}%
In order to measure $X_{a}$, the probe observable is chosen best to be 
$M=X_{b}/G$. 
Then we have
\beq
M^{\Out}=X_{a}^{\In}+\frac{1}{G}X_{b}^{\In}.
\eeq
The $X_{a}$-noise operator, 
the $X_{a}$-disturbance operator,
and the $Y_{a}$-disturbance operator
are given by
\begin{mathletters}
\beqa
N(X_{a})&=&\frac{1}{G}X_{b}^{\In},\\
D(X_{a})&=&0,\\
D(Y_{a})&=&-GY_{b}^{\In}.
\eeqa
\end{mathletters}%
The condition $D(X_{a})=0$ is characteristic of BAE amplifiers.
If the probe is prepared nearly in the $X_{b}$ eigenstate
$\ket{X_{b}=0}$, the measurement is a nearly noiseless ($\ep(X_{a})\approx 0$)
and nondisturbing ($\et(X_{a})=0$) measurement of $X_{a}$. 
From the above, BAE amplifiers are of independent intervention
for the pair $(X_{a},Y_{a})$.
Thus, the Heisenberg noise-disturbance uncertainty relation for 
the pair $(X_{a},Y_{a})$ holds.  
If the $b$-mode is prepared in the vacuum, $\xi=\ket{0}$,
the noise and disturbance satisfy
$\ep(X_{a})=1/2$ and 
$\et(Y_{a})=1/2$, so that the minimum noise-disturbance 
product attains as
$\ep(X_{a})\et(Y_{a})=1/4$.

\section{Noiseless quadrature transducers}

Consider the following input-output relations
\begin{mathletters}\label{eq:712a}
\beqa
X_{a}^{\Out}&=&X_{a}^{\In}-X_{b}^{\In},\label{eq:711g}\\
X_{b}^{\Out}&=&X_{a}^{\In},\label{eq:711i}\\
Y_{a}^{\Out}&=&-Y_{b}^{\In},\label{eq:711h}\\
Y_{b}^{\Out}&=&Y_{b}^{\In}+Y_{a}^{\In}.\label{eq:711j}
\eeqa
\end{mathletters}%
In this case, the measuring interaction $U$ is called the {\em noiseless
quadrature transducer}.
In order to measure $X_{a}$, the probe observable is chosen to be 
$M=X_{b}$. 
The $X_{a}$-noise operator, the $X_{a}$-disturbance operator, 
and the $Y_{a}$-disturbance operator are given by
\begin{mathletters}
\beqa
N(X_{a})&=&0,\label{eq:711d}\\
D(X_{a})&=&-X_{b}^{\In},\label{eq:711e}\\
D(Y_{a})&=&-(Y_{a}^{\In}+Y_{b}^{\In}).\label{eq:711f}
\eeqa
\end{mathletters}%
The condition $N(X_{a})=0$ is characteristic of the noiseless transducer.
Hence, the measurement is a noiseless $X_{a}$-measurement regardless
of the probe preparation $\xi$.
From \eq{711e},
if the probe is prepared nearly in the $X_{b}$ eigenstate
$\ket{X_{b}=0}$, the measurement is a noiseless ($\ep(X_{a})=0$) and nearly 
nondisturbing ($\et(X_{a})\approx 0$) measurement of $X_{a}$. 
Since $\ep(X_{a})=0$, we have
\beq
\ep(X_{a})\et(Y_{a})=0
\eeq
for any states $\ps$ and $\xi$, so that Heisenberg's lower bound for
the noise-disturbance product can be overcome by a noiseless and nearly
nondisturbing quadrature measurement,
if one can implement a noiseless quadrature transducer.
The above model also suggests that the linearity of measuring interaction
does not ensure the validity
of the Heisenberg noise-disturbance uncertainty relation, despite a 
prevailing claim that linear measurements, measurements closely 
connected to linear systems, obey the Heisenberg noise-disturbance
uncertainty relation${\cite{BK92}}$.

\section{Experimental realization of the noiseless quadrature transducers}

The noiseless quadrature transducer can be implemented, in principle, as follows.
Consider two degenerate modes $a$, the signal, 
and $b$, the probe, with frequency $\om$ and
orthogonal polarization, which undergo successive parametric 
interactions in the following 5 steps;
see Refs.~\cite{Yur85,PSY89,SCY90,POK94,GLP98} for similar
implementations of BAE amplifiers. 

(i) The two polarization modes undergo a mixing interaction using 
a polarization rotator which rotates the angle of polarization by
$\th$. The operation of the polarization rotator is 
represented by the mixing operator 
\beq
T(\th)=\exp[\th(ab^{\da}-a^{\da}b)].
\eeq

(ii) The mixture of the signal and the probe fields will propagate along
each of the ordinary and orthogonal extraordinary axis of a KTP crystal
pumped by a pulsed intense classical field.
This interaction is a nondegenerate parametric amplifier described by 
the two-mode squeeze operator
\beq
S(r)=\exp[r(ab-a^{\da}b^{\da})],
\eeq
where $r$ corresponds to the squeezing parameter.

(iii) After the amplification step the fields pass through a second
polarization rotator with the mixing angle $2\th$ so that the
operation is represented by $T(2\th)=T(\th)^{2}$.

(iv) The mixture of the signal and the probe fields undergoes the
second parametric amplification described by the two-mode
squeeze operator $S(-r)$.

(v) After the second amplification step the field pass through 
a third polarization rotator with the mixing angle $\th$ so that
the operation is represented by $T(\th)$.

Thus this process defines the measuring interaction operator
\beq
U(r,\th)=T(\th)S(-r)T(\th)T(\th)S(r)T(\th).
\eeq
We shall determine the parameters $\th$ and $r$ for $U(r,\th)$ 
to realize a noiseless quadrature transducer, \Eq{712a}.
 Suppose that $\th$ and $r$ satisfies the relation $\sin 2\th=
\tanh r$.  In this case, it is well-known${\cite{Yur85,PSY89,SCY90,POK94,GLP98}}$
that the unitary  operator
$U_{-}=T(\th)S(-r)T(\th)$ realizes the BAE quadrature amplifier with
$G=2\sinh r$, i.e.,
\begin{mathletters} 
\beqa
U_{-}^{\da}X_{a}U_{-}&=&X_{a},\\
U_{-}^{\da}X_{b}U_{-}&=&X_{b}+2(\sinh r)X_{a},\\
U_{-}^{\da}Y_{a}U_{-}&=&Y_{a}-2(\sinh r)Y_{b},\\
U_{-}^{\da}Y_{b}U_{-}&=&Y_{b}.
\eeqa
\end{mathletters}%
Similarly, the unitary operator $U_{+}=T(\th)S(r)T(\th)$
realizes the conjugate BAE quadrature amplifier, i.e.,
\begin{mathletters}
\beqa
U_{+}^{\da}X_{a}U_{+}&=&X_{a}-2(\sinh r)X_{b},\\
U_{+}^{\da}X_{b}U_{+}&=&X_{b},\\
U_{+}^{\da}Y_{a}U_{+}&=&Y_{a},\\
U_{+}^{\da}Y_{b}U_{+}&=&Y_{b}+2(\sinh r)Y_{a}.
\eeqa
\end{mathletters}%
Combining the above equations, we have the input-output
relations for the unitary operator $U(r,\th)=U_{-}U_{+}$,
\begin{mathletters}
\beqa
U(r,\th)^{\da}X_{a}U(r,\th)&=&X_{a}-2(\sinh r)X_{b},\\
U(r,\th)^{\da}X_{b}U(r,\th)&=&(1-4\sinh^{2}r)X_{b}+2(\sinh r)X_{a},\\
U(r,\th)^{\da}Y_{a}U(r,\th)&=&(1-4\sinh^{2}r)Y_{a}-2(\sinh r)Y_{b},\\ 
U(r,\th)^{\da}Y_{b}U(r,\th)&=&Y_{b}+2(\sinh r)Y_{a}.
\eeqa
\end{mathletters}%
Thus, if $\sinh r=1/2$, i.e., 
\begin{mathletters}\label{eq:choice}
\beqa
r&=&\log \frac{1+\sqrt{5}}{2}\\
\th&=&\frac{1}{2}\sin^{-1}\frac{1}{\sqrt{5}},
\eeqa
\end{mathletters}%
the resulting unitary operator $U=U(r,\th)$ realizes the
noiseless quadrature transducer, \Eq{712a}.

Thus, if $r$ and $\th$ are chosen as \Eq{choice}, 
the parametric process (i)--(v) realizes the noiseless quadrature transducer.
Therefore, this process followed immediately by the homodyne detection
of the $X_{b}$ component implements a nearly nondisturbing and
noiseless measurement of the $X_{a}$ quadrature component
that disturbs the conjugate observable $Y_{a}$ much less than the quantum 
limit set by Heisenberg's lower bound for the noise-disturbance product,
$\ep(X_{a})\et(Y_{a})\ge 1/4$.

\section{Conclusion}

In this paper, I have proposed new relations, 
\eq{UVUR} and  \eq{GNDUR}, universally
valid for the trade-off between the measurement noise and disturbance.
These relations demonstrates that the prevailing Heisenberg's lower bound 
for the noise-disturbance product is valid for measurements with
independent intervention, but can be circumvented by a measurement with
dependent intervention. An experimental confirmation of the violation of
Heisenberg's lower bound is proposed for a measurement of optical
quadrature with currently available techniques in quantum optics. 
The new relation will not only bring a new insight on fundamental 
limitations on measurements set by quantum mechanics but also advance a
frontier of precision measurement technology such as gravitational wave
detection and quantum information processing.

\acknowledgments
{\bf Acknowledgments.} 
The author thanks F. De Martini, S. Takeuchi,
and I. Ojima for useful discussions. 
This work was supported by the programme ``R\&D on Quantum  
Communication Technology'' of the MPHPT of Japan, by the CREST project of the
JST, and by the Grant-in-Aid for Scientific Research of the JSPS.

\end{document}